\documentclass[reprint,prl,aps]{revtex4-1}
\usepackage{graphicx}
\usepackage{color}
\usepackage{dcolumn}
\usepackage{bm}
\usepackage{amssymb}
\usepackage{color,amsmath}
\usepackage{soul,xcolor} %use \st{text} to achieve strikethrough
\setstcolor{red} %set colour of strikethrough to red
\usepackage{epstopdf}

\newcolumntype{C}[1]{>{\centering\arraybackslash}m{#1}}
\newcolumntype{N}{@{}m{0pt}@{}}

\definecolor{cadmiumgreen}{rgb}{0.0, 0.42, 0.24}

\begin{document}
% \linenumbers

\title{Tunable correlation-driven symmetry breaking in twisted double bilayer graphene}

\author{Minhao He$^{1}$} 
\author{Yuhao Li$^{1}$} 
\author{Jiaqi Cai$^{1}$}
\author{Yang Liu$^{1}$} 
\author{K. Watanabe$^{2}$} 
\author{T. Taniguchi$^{2}$} 
\author{Xiaodong Xu$^{1,3\dagger}$}  
\author{Matthew Yankowitz$^{1,3\dagger}$}

\affiliation{$^{1}$Department of Physics, University of Washington, Seattle, Washington, 98195, USA}
\affiliation{$^{2}$National Institute for Materials Science, 1-1 Namiki, Tsukuba 305-0044, Japan}
\affiliation{$^{3}$Department of Materials Science and Engineering, University of Washington, Seattle, Washington, 98195, USA}
\affiliation{$^{\dagger}$ xuxd@uw.edu (X.X.); myank@uw.edu (M.Y.)}

\maketitle

\textbf{A variety of correlated phases have recently emerged in select twisted van der Waals (vdW) heterostructures owing to their flat electronic dispersions~\cite{Cao2018a,Cao2018b,Yankowitz2019,Sharpe2019,Cao2019,Polshyn2019,Lu2019,Serlin2019,Stepanov2019,Saito2019,Shen2019,Liu2019,Cao2019b,Burg2019,Chen2019a,Chen2019b,Chen2019c,Tang2019,Regan2019,Wang2019b}. In particular, heterostructures of twisted double bilayer graphene (tDBG) manifest electric field-tunable correlated insulating (CI) states at all quarter fillings of the conduction band, accompanied by nearby states featuring signatures suggestive of superconductivity~\cite{Shen2019,Liu2019,Cao2019b,Burg2019}. Here, we report electrical transport measurements of tDBG in which we elucidate the fundamental role of spontaneous symmetry breaking within its correlated phase diagram. We observe abrupt resistivity drops upon lowering the temperature in the correlated metallic phases neighboring the CI states, along with associated nonlinear $I$-$V$ characteristics. Despite qualitative similarities to superconductivity, concomitant reversals in the sign of the Hall coefficient instead point to spontaneous symmetry breaking as the origin of the abrupt resistivity drops, while Joule heating appears to underlie the nonlinear transport. Our results suggest that similar mechanisms are likely relevant across a broader class of semiconducting flat band vdW heterostructures.}

Materials with small electronic bandwidths typically manifest effects of electronic correlations at low temperature owing to the dominant role of Coulomb interactions between charge carriers. Recently, a new class of correlated materials has arisen in select heterostructures of atomically-thin van der Waals crystals with a moir\'e superlattice, and can be grouped into two categories based on the band structures of their parent materials. For example, twisted bilayer graphene (tBLG) consists of two semimetallic monolayer graphene building blocks, and Dirac crossings in tBLG are protected by the product of two-fold rotation and time reversal symmetry ($C_2 T$)~\cite{Po2018}. The band structure of tBLG is determined by a complicated interplay of moir\'e-mediated interlayer tunneling parameters~\cite{Bistritzer2011}, with uniquely flat low energy bands emerging over an extremely narrow range of twist angles very near 1.1$^{\circ}$~\cite{Cao2018a,Cao2018b,Yankowitz2019,Sharpe2019,Serlin2019,Cao2019,Polshyn2019,Lu2019,Stepanov2019,Saito2019,Jiang2019,Kerelsky2019,Xie2019,Choi2019}. In contrast, all other correlated flat band moir\'e vdW heterostructures investigated so far consist of semiconducting parent crystals – including materials such as WSe$_2$ and WS$_2$ which inherently have wide band gaps~\cite{Tang2019,Regan2019,Wang2019b}, and materials such as bilayer graphene and ABC trilayer graphene in which band gaps can be induced by applying a perpendicular electrical displacement field, $D$~\cite{Shen2019,Liu2019,Cao2019b,Burg2019,Adak2020,Chen2019a,Chen2019b,Chen2019c}. Flat bands emerge over a wider range of twist angles in these systems owing to the gap at the charge neutrality point (CNP), and can be further modified with applied $D$.

Twisted double bilayer graphene --- comprising two sheets of Bernal-stacked bilayer graphene rotated by an angle $\theta$ --- provides a model platform for investigating the properties of a semiconducting flat band moir\'e vdW heterostructure. Previous studies of tDBG have identified a robust correlated insulating state at half filling of the conduction band stabilized over a finite range of $D$, as well as additional CI states at one- and three-quarters fillings which emerge in a magnetic field~\cite{Shen2019,Liu2019,Cao2019b,Burg2019}. Abrupt drops in the resistivity of the neighboring metallic states are observed as the temperature is lowered, saturating to small values at low temperature~\cite{Shen2019,Liu2019,Cao2019b}. In addition, these states exhibit highly nonlinear $I$-$V$ characteristics~\cite{Liu2019}. Together, this raises the possibility that these transport features arise as a consequence of superconductivity~\cite{Shen2019,Liu2019}. Although anomalous transport features in the valence band have been reported as well, the role of correlations in these states remains unclear. Overall, transport in tDBG is complicated as it is determined by a number of factors which appear to be of similar magnitude: width of the flat bands, energy gaps isolating these bands, and strength of Coulomb interactions. So far, a complete understanding of the correlated phase diagram of tDBG remains lacking.

%%%%%%%%%%%%%%%%%%%%%%%%%%%%%%%%%%%%%%%%%%%%%%%%%%%%%%%%%%%%%%%%%%%%%%
\begin{figure*}[t]
\includegraphics[width=6.9 in]{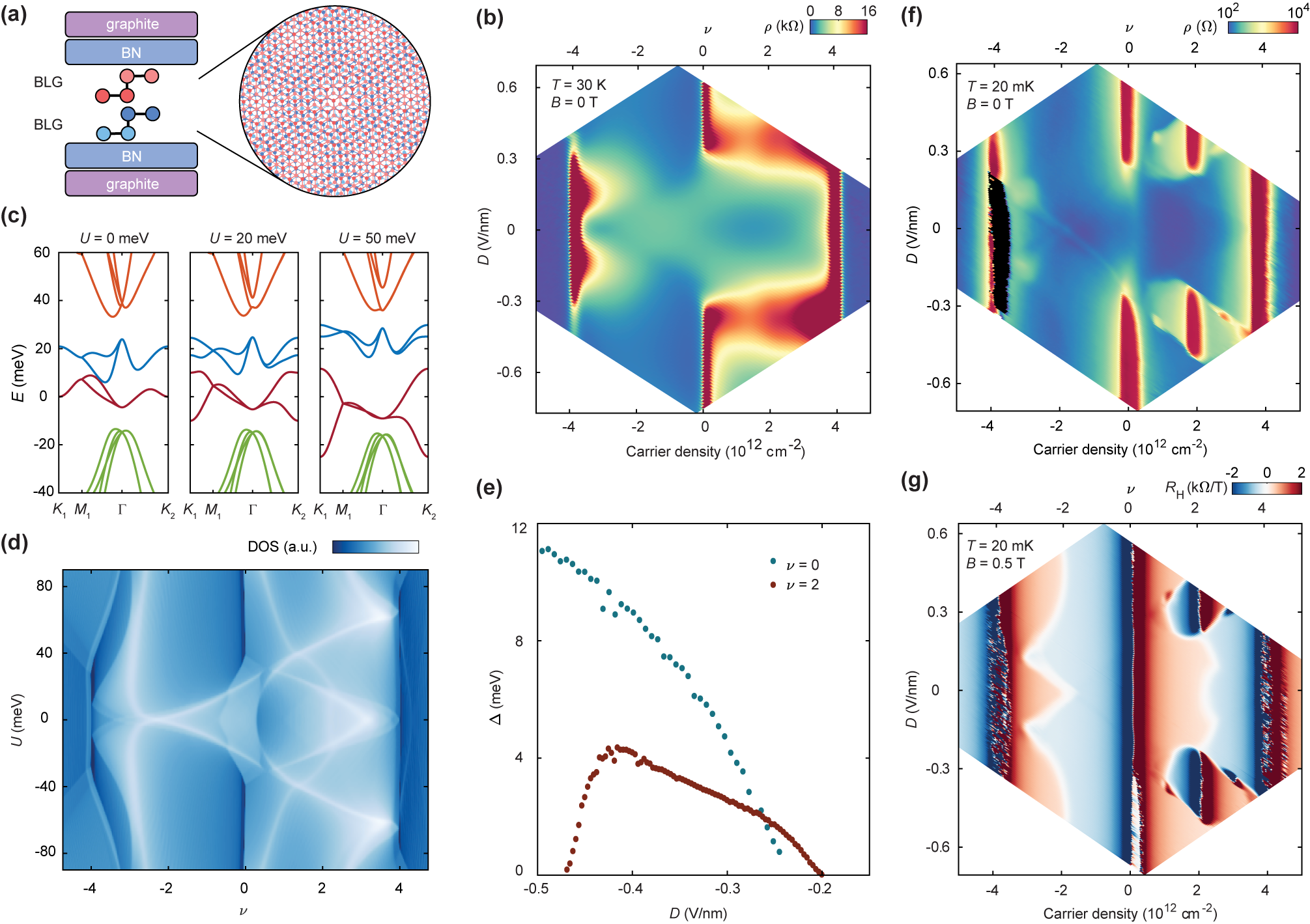} 
\caption{\textbf{Tunable band structure and transport in a 1.30$^{\circ}$ tDBG device.}
\textbf{a}, Schematic of a tDBG device encapsulated in BN with graphite gates.
\textbf{b}, Resistivity of a tDBG device with $\theta = 1.30^{\circ}$ at $T = 30$~K. The corresponding band filling factor $\nu$ is shown on the top axis.
\textbf{c}, Calculated band structure of tDBG with $\theta = 1.30^{\circ}$ at various values of interlayer potential, $U$.
\textbf{d}, Calculated DOS as a function of $\nu$ and $U$, plotted on a log color scale. 
\textbf{e}, Energy gaps of the CNP ($\nu = 0$) and the CI state at $\nu = 2$ as a function of $D$ measured by thermal activation.
\textbf{f}, Resistivity at $T = 20$~mK. The region colored in black corresponds to artificially negative resistance in the contacts used for this measurement.
\textbf{g}, Hall coefficient, $R_H$, antisymmetrized at $|B| = 0.5$~T.
}
\label{fig:1}
\end{figure*}
%%%%%%%%%%%%%%%%%%%%%%%%%%%%%%%%%%%%%%%%%%%%%%%%%%%%%%%%%%%%%%%%%%%%%%

Here, we investigate temperature-dependent electrical transport of dual-gated tDBG (Fig.~\ref{fig:1}a) in four devices with $\theta$ ranging from 1.17$^{\circ}$ to 1.53$^{\circ}$. We restrict our attention to a device with $\theta = 1.30^{\circ}$, although we have observed qualitatively similar behavior in all devices (see Supplementary Section 1 and Supplementary Fig.~\ref{fig:S2}). We first identify transport features corresponding to the single-particle band structure of tDBG by measuring transport at high temperature, such that correlated states are suppressed. Figure~\ref{fig:1}b shows the device resistivity, $\rho$, as a function of charge density, $n$, and $D$ at temperature $T = 30$~K. The top axis denotes the filling factor of the bands, $\nu$ (see Methods). For small $|D| \lesssim 0.3$~V/nm, the device exhibits only weak resistivity modulations as $n$ and $D$ are tuned. Most notably, a broad cross-like feature is observed for hole-type doping ($\nu < 0$). For larger $|D|$, we observe a substantial difference in $\rho$ between electron- and hole-type doping.

We compare our observations to the single-particle band structure of tDBG, calculated following the procedure developed in Ref.~\cite{Lee2019} with $\theta = 1.30^{\circ}$. Figure~\ref{fig:1}c shows the band structure for various values of the interlayer potential, $U$, and Fig.~\ref{fig:1}d plots the corresponding density of states (DOS) as a function of $\nu$ and $U$. The model predicts a semimetal-to-semiconductor transition at the CNP ($\nu$ = 0) with increasing $|D|$, as well as band gaps at full filling of the valence and conduction bands ($\nu$ = $\pm$4, where the factor of 4 accounts for the spin and valley degeneracy of the bands) which diminish with increasing $|D|$. These features correspond well with insulating states observed over similar ranges of $|D|$ in our measurement (Fig.~\ref{fig:1}b). Additionally, the model predicts a bifurcation of the van Hove singularity in the valence band with increasing $|D|$, forming a cross-like feature in the DOS (white contours in Fig.~\ref{fig:1}d for $\nu < 0$) reminiscent of our observed cross-like transport feature in Fig.~\ref{fig:1}b. Finally, the model predicts that the conduction band becomes flatter with increasing $|D|$, whereas the valence band becomes more dispersive (Fig.~\ref{fig:1}c). When the two bands are isolated by a gap at the CNP (\textit{i.e.} at large $|D|$), differences in mobility arising from the relative flatness of each band are likely responsible for the large discrepancy in $\rho$ observed between electron- and hole-type doping. Thermal activation measurements at $\nu =0$ confirm that a gap emerges for $|D| \gtrsim 0.25$~V/nm and grows monotonically upon further increasing $|D|$ (Fig.~\ref{fig:1}e).

We now turn our attention to transport at our base temperature of $T = 20$~mK, shown in Fig.~\ref{fig:1}f. Most of the transport features described above persist to base temperature, however we observe the additional emergence of new features in the conduction band at large $|D|$ which are not anticipated within the single-particle model. A well developed insulating state emerges at half filling of the conduction band ($\nu = 2$) over a finite range of $|D|$, as well as incipient insulating states at $\nu = 1$ and 3. We additionally observe tilted halo-like features surrounding these states characterized by a weak enhancement of $\rho$. These states arise as a consequence of correlations, as previously identified in prior studies of tDBG~\cite{Shen2019,Liu2019,Cao2019b,Burg2019}. Thermal activation measurements of the energy gap of the $\nu = 2$ CI state mark the approximate range of $|D|$ over which correlated states are observed. In particular, CI states emerge when $|D|$ is approximately large enough to gap the bands at $\nu = 0$. The suppression of the CI gap at even larger $|D| \approx 0.47$~V/nm is consistent with prior reports, in which this was argued to be tied to the gap closing at $\nu = 4$~\cite{Shen2019,Liu2019,Cao2019b,Burg2019}. However, we are not able to directly confirm this in our device owing to limitations in the accessible range of gate voltage.

Corresponding measurements of the Hall coefficient, $R_H = (R_{xy}[B] - R_{xy}[-B])/(2B)$, provide additional insight into the nature of these correlated states (Fig.~\ref{fig:1}g). The small applied $|B| = 0.5$~T does not substantially alter the transport (Supplementary Fig.~\ref{fig:S5}), and can therefore be directly compared with Fig.~\ref{fig:1}f. We observe rapid sign changes in $R_H$ as a function of carrier density at nearly all of the insulating states ($\nu = 0$, 2, 3, and $\pm$4), as well as a weaker modulation near the incipient CI state at $\nu = 1$, owing to the sign change of the carrier mass on opposite sides of the band gaps. $R_H$ also changes sign at partial band filling in regions of the phase diagram that are well described by single-particle considerations, consistent with the required reversal of the carrier sign as the band is tuned from empty to filled. We observe additional sign changes in $R_H$ corresponding nearly exactly to the boundary of the halo observed in measurements of $\rho$ in Fig.~\ref{fig:1}f. Furthermore, replica halo features are also apparent in $R_H$ surrounding $\nu = 1$ and 3. As we will discuss in detail later, these sign reversals arise as a consequence of correlations restructuring the bands.

%%%%%%%%%%%%%%%%%%%%%%%%%%%%%%%%%%%%%%%%%%%%%%%%%%%%%%%%%%%%%%%%%%%%%%
\begin{figure*}[t]
\includegraphics[width=6.9 in]{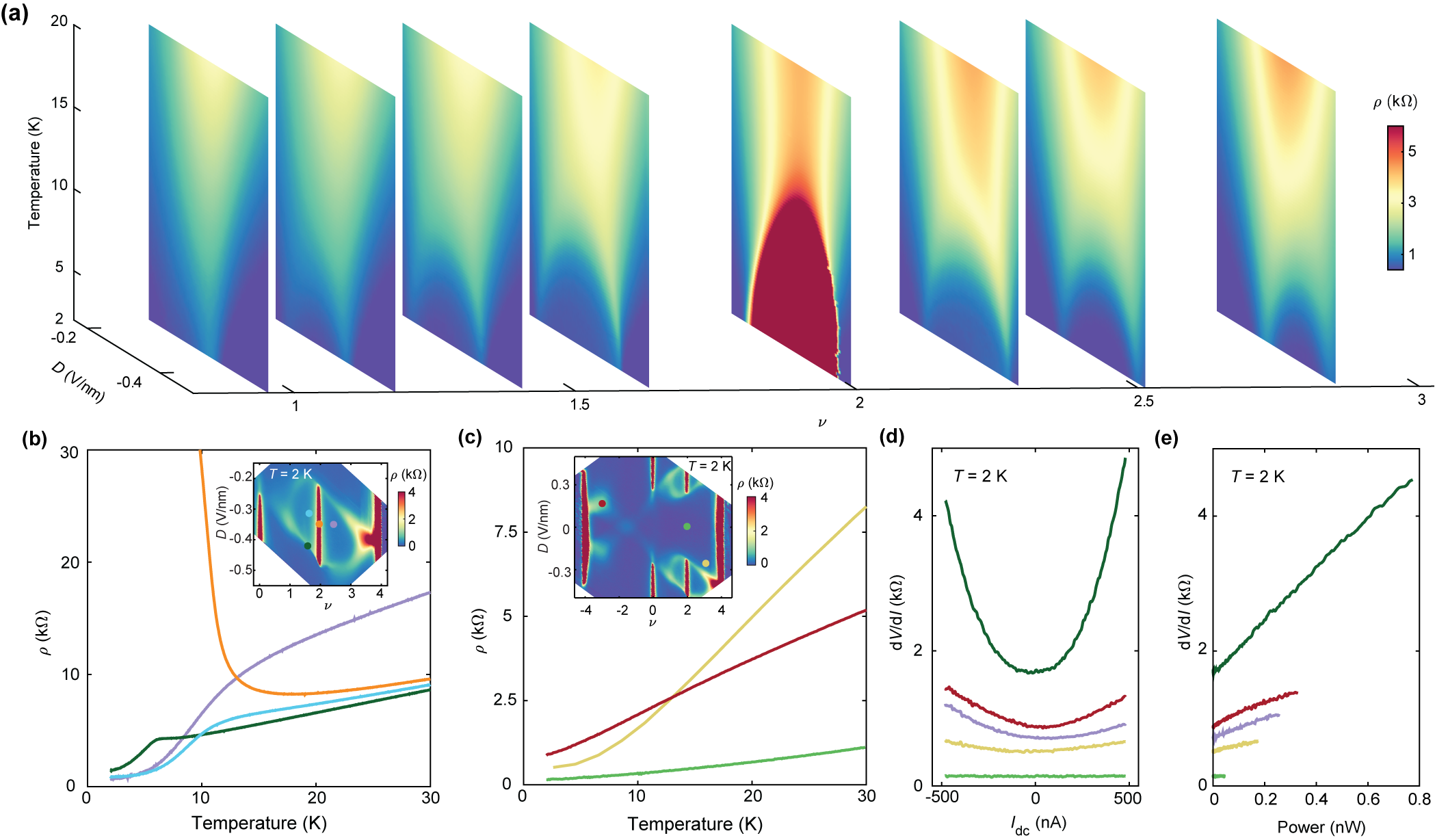} 
\caption{\textbf{Temperature-dependent transport and Joule heating in tDBG.}
\textbf{a}, Resistivity as a function of temperature and displacement field at a few select values of $\nu$.  
\textbf{b, c}, $\rho(T)$ inside and outside the halo region, respectively, at select values of $(\nu,D)$ denoted by the color-coded circle markers in the insets. (Insets) Resistivity at $T = 2$~K.
\textbf{d}, Differential resistance, d$V$/d$I$ versus applied dc current bias, $I_{dc}$, at $T = 2$~K, with colors corresponding to the curves in \textbf{b}-\textbf{c}.
\textbf{e}, d$V$/d$I$ as a function of the calculated Joule heating power, $P = IV$.
}
\label{fig:2}
\end{figure*}
%%%%%%%%%%%%%%%%%%%%%%%%%%%%%%%%%%%%%%%%%%%%%%%%%%%%%%%%%%%%%%%%%%%%%%

We investigate the properties of the metallic states by measuring the temperature-dependence of $\rho$. Figure~\ref{fig:2}a shows $\rho(T)$ as a function of $D$ for a few select values of $\nu$. The development of the CI state at $\nu = 2$ results in an arch-like feature centered at optimal $D$, with an associated metal-insulator transition at $T \approx 15$~K. The arch-like feature persists for $\nu \neq 2$, however in contrast $\rho$ continues to decrease monotonically as the temperature is lowered. The maximum onset temperature of the arch decreases as $\nu$ is detuned further from 2, with the arms of the arches at base temperature corresponding to the boundaries of the primary halo in Figs.~\ref{fig:1}f-g. 

Figure~\ref{fig:2}b shows $\rho(T)$ at four values of $(\nu,D)$ within the halo, as denoted in the figure inset. The orange curve exhibits a metal-insulator transition associated with the development of the CI state at $\nu = 2$. Away from $\nu = 2$, $\rho(T)$ scales linearly with $T$ at high temperatures, but further exhibits an abrupt drop at a temperature corresponding to the formation of the arch in Fig.~\ref{fig:2}a, before eventually beginning to saturate to a finite residual resistivity at low $T$. The dark green curve, acquired precisely on the halo boundary, exhibits the most rapid drop in $\rho(T)$. In contrast, transport outside the halo region does not manifest such resistivity drops, instead always exhibiting linear-in-$T$ resistivity until eventually saturating to a finite residual resistivity at low $T$ (Fig.~\ref{fig:2}c). Such behavior is commonly observed in graphene~\cite{Chen2008,Dean2010}, however here we observe a very large slope of a few hundred ohms/Kelvin comparable to prior observations in tBLG~\cite{Cao2019,Polshyn2019}.

%%%%%%%%%%%%%%%%%%%%%%%%%%%%%%%%%%%%%%%%%%%%%%%%%%%%%%%%%%%%%%%%%%%%%%
\begin{figure*}[t]
\includegraphics[width=6.2 in]{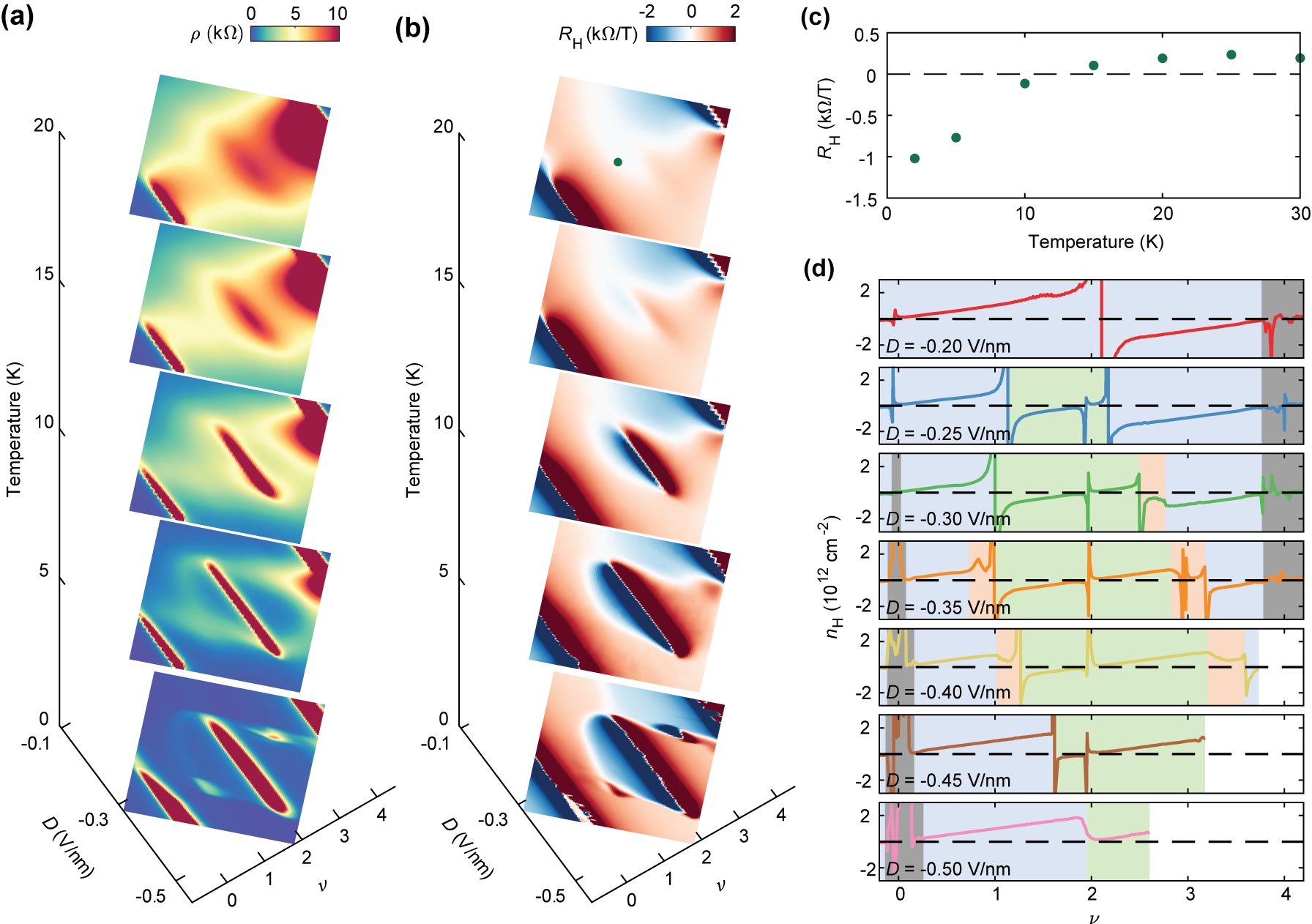} 
\caption{\textbf{Evidence for spontaneous symmetry inside the halo region.}
\textbf{a}, Resistivity acquired at $T = 20$~K, 15 K, 10 K, 5 K and 20 mK, respectively (top to bottom).  
\textbf{b}, Corresponding maps of the Hall coefficient, $R_H$.
\textbf{c}, Hall coefficient, $R_H$, as a function of $T$ acquired at slight underdoping of the CI state at $\nu = 2$ (corresponding to the green circle marker in \textbf{b}). 
\textbf{d}, Hall density, $n_H$, as a function of $\nu$ at various $D$. Color shading denotes regions of the phase diagram with different inferred degeneracy, where blue corresponds to four-fold degeneracy, green corresponds to two-fold degeneracy, and orange has no remaining degeneracy. Gray-shaded regions denote insulating states at $\nu = 0$ and 4.
}
\label{fig:3}
\end{figure*}
%%%%%%%%%%%%%%%%%%%%%%%%%%%%%%%%%%%%%%%%%%%%%%%%%%%%%%%%%%%%%%%%%%%%%%

The abrupt drops in $\rho(T)$ in curves acquired within the halo region resemble the onset of superconductivity. However, despite heavy electronic filtering, we observe a finite residual resistivity at base temperature of a few hundred ohms in all devices. As an additional check, we measure the differential resistance, d$V$/d$I$, as a function of applied dc current bias, $I_{dc}$ at $T = 2$~K (Fig.~\ref{fig:2}d). The curves are acquired at the same $(\nu,D)$ as in Figs.~\ref{fig:2}b-c. Inside the halo (dark green and purple curves), we observe nonlinear transport that is not generally anticipated for a typical metal. Although superconducting states also exhibit nonlinear $I$-$V$, we do not observe signatures of a well defined critical current in our measurements, nor do we observe saturation to a ``normal state'' value of d$V$/d$I$ up to $I_{dc} = 500$~nA. Furthermore, nonlinear transport is not limited solely to curves acquired within the halo, but is also observed outside the halo in cases where the corresponding $\rho(T)$ does not exhibit an abrupt drop (yellow and red curves). Figure~\ref{fig:2}e shows d$V$/d$I$ as a function of the dissipated power $P = IV$. In general, larger nonlinearities in $I$-$V$ correspond to higher power dissipation, pointing to Joule heating as the most likely underlying mechanism. In contrast, transport is linear in the light green curve, which exhibits the lowest $\rho$ at all $T$. Nonlinear transport in our devices is generically observed in states with large residual resistivity at base temperature and large d$\rho$/d$T$, and therefore does not appear to be a signature of superconductivity. 

Figures~\ref{fig:3}a-b show maps of $\rho$ and $R_H$, respectively, surrounding the halo region from $T = 20$~mK to 20 K acquired in $\sim$5 K steps. The CI state and the surrounding resistive halo smear together as the temperature is raised (Fig.~\ref{fig:3}a), eventually forming a highly resistive stripe spanning the entire conduction band. In the corresponding $R_H$ (Fig.~\ref{fig:3}b), the $R_H = 0$ contour shrinks towards $\nu = 2$ as the temperature is raised. This is particularly evident for $\nu < 2$, as the corresponding contour for $\nu > 2$ is complicated by thermal activation to the higher-energy remote band~\cite{Polshyn2019}. Within much of the halo region, we find that the sign of $R_H$ reverses as the temperature is lowered (Fig.~\ref{fig:3}c, corresponding to the green circle marker in Fig.~\ref{fig:3}b).

Given the apparent association of the halo features in $\rho$ with the sign changes in $n_H$, we interpret the halo regions as footprints of spontaneous symmetry breaking within the phase diagram of tDBG. Figure~\ref{fig:3}d shows the corresponding evolution of the Hall density, $n_H = e/R_H$, for select values of $D$ at $T = 20$~mK. While $n_H$ exhibits a single sign change near $\nu = 2$ at small $D$, the emergence of the CI states results in additional sign changes and abrupt resets of $n_H$ at integer $\nu$. The sign of $n_H$ follows single-particle expectations within the blue-shaded regions of Fig.~\ref{fig:3}d. We therefore infer that these regions correspond to metallic states with full spin and valley degeneracy. The green shading corresponds to the primary halo region observed in Figs.~\ref{fig:3}a-b, and is bounded by additional sign changes in $n_H$. This denotes regions of the phase diagram associated with the formation of the CI state at $\nu = 2$, in which we presume that a single degeneracy is lifted. The orange shading corresponds to the replica halos in Figs.~\ref{fig:3}a-b, and is bounded by additional abrupt resets in $n_H$. This marks regions of the phase diagram which we presume to have no remaining degeneracies owing to the additional development of the CI states at $\nu = 1$ and 3.

Collectively, our results suggest that correlations dramatically restructure the Fermi surface as a function of $n$, $D$, and $T$. In particular, correlated states emerge as $T$ is lowered, and appear to be strongest at optimal values of $\nu \approx 2$ and $D \approx 0.35$~V/nm at this twist angle. At base temperature, detuning $n$ and $D$ from these values further modifies the band structure and results in the formation of additional correlated ground states. Within the halo region, abrupt drops in $\rho(T)$ arise concomitantly with sign changes in $R_H$, and therefore appear to be directly associated with the development of symmetry-broken states within the isolated flat band.

Our measurements of the energy gap of the $\nu = 2$ CI state do not scale trivially with in-plane magnetic field (see Supplementary Section 2 and Supplementary Fig.~\ref{fig:S6}), possibly due to a unique orbital contribution in tDBG~\cite{Lee2019} which arises in addition to the usual Zeeman energy. Although we are not able to unambiguously determine the ground state ordering of the CI states or the neighboring symmetry-broken metallic states, the CI states always become more resistive with in-plane field suggestive of ferromagnetic ordering. The precise mechanism driving the abrupt drop in $\rho(T)$ within the halo region in tDBG remains an open question, however qualitatively similar transport has previously been observed in certain bulk crystals owing to reduced scattering associated with the emergence of various types of magnetic ordering~\cite{Kasuya1956}, including in MnSi~\cite{Petrova2006} and BaFe$_2$As$_2$~\cite{Rotter2008}. Although we do not observe any signatures of an anomalous Hall effect within the halo region (Supplementary Fig.~\ref{fig:S7}), numerous magnetically-ordered ground states with net zero Berry curvature remain possible even in its absence. The abrupt reduction in $\rho$ may also be associated with reduced inelastic scattering in the symmetry-broken band or from variations in the strength of electron-phonon coupling. Further theoretical analysis is necessary to provide a detailed quantitative understanding of these observations.

Finally, it is interesting to consider our results in the context of other semiconducting flat band moir\'e vdW heterostructures. Recent studies of twisted WSe$_2$~\cite{Wang2019b} and ABC trilayer graphene aligned with BN~\cite{Chen2019b} have also reported displacement field-tunable CI states emerging at integer $\nu$, along with associated reversals in the sign of $n_H$ on either side of the CI states. Similar to tDBG, CI states emerge in trilayer graphene only at large $|D|$, when a gap at the CNP isolates the bands. When the combination of $(\nu,D)$ is detuned slightly from the CI states, abrupt drops in $\rho(T)$ and associated nonlinear $I$-$V$ suggestive of superconductivity are also observed~\cite{Chen2019b,Wang2019b}. Often, the residual resistivity at base temperature is not precisely zero~\cite{Chen2019b}. Many of these observations contrast with tBLG, in which unambiguous signatures of superconductivity are routinely observed (Supplementary Section 4). Although direct comparisons are difficult, the qualitative similarities of our results with the other semiconducting flat band systems studied so far motivates the possibility that in all such cases, the nonlinear $I$-$V$ may be governed by Joule heating and the abrupt resistivity drops may result from spontaneous symmetry breaking.

\section*{Methods}
tDBG devices are fabricated using the ``tear-and-stack'' method~\cite{Kim2016,Cao2016}, or by first cutting the bilayer graphene using an atomic force microscope tip in order mitigate strain in the heterostructure~\cite{Chen2019a,Serlin2019}. tDBG is encapsulated between flakes of hexagonal boron nitride (BN) with typical thickness of 30 - 50 nm. The devices feature graphite top and bottom gates with typical flake thicknesses around 5 nm. Heterostructures of graphite/BN/tDBG/BN/graphite are assembled using standard dry-transfer techniques with a polycarbonate film on top of a polydimethyl siloxane stamp~\cite{Wang2013}. Completed heterostructures are transferred onto a Si/SiO$_2$ wafer. Conventional plasma etching and metal deposition techniques are utilized to fabricate samples into Hall bar geometries~\cite{Wang2013}.

Transport measurements are conducted in a four-terminal geometry with typical a.c. current excitations of 5–10 nA using a standard lock-in technique at 13.3 Hz. Regions of the tDBG extend beyond both the graphite top and bottom gates, and in some measurements are doped by applying a voltage to the Si gate. Transport characterization at temperatures above 2 K was performed in a PPMS DynaCool. Transport characterization down to 20 mK was performed in a Bluefors dilution refrigerator with low temperature electronic filtering. 

The twist angle $\theta$ is determined from the values of $n$ at which the insulating states at full band filling ($\nu = \pm$4) appear, following $n = 8\theta^2/\sqrt{3}a^2$, where $a$ = 0.246 nm is the lattice constant of graphene. The filling factor is defined as $\nu = \sqrt{3}\lambda^2 n/2$, where $\lambda$ is the period of the moir\'e. The charge carrier density corresponding to $\nu = \pm$4 is determined by tracing the sequence of quantum oscillations in a magnetic field projecting to full band filling at $B$ = 0. 

The dual-gated geometry of the devices allows independent tuning of $n$ and $D$. We calculate $n = (V_tC_t + V_bC_b)/e$, and $D = (V_tC_t – V_bC_b)/2\epsilon_0$, where $C_t$ and $C_b$ are the top and bottom gate capacitances, $V_t$ and $V_b$ are the top and bottom gate voltages, $e$ is the electron charge, and $\epsilon_0$ is the vacuum permittivity.

The band structure of tDBG is calculated using a generalization of Bistritzer-MacDonald model~\cite{Bistritzer2011}. The effect of the moir\'e potential is captured by a tight-binding model in an effective $K$-space honeycomb lattice following the details of Ref.~\cite{Lee2019}, in which the effects of lattice relaxations are captured phenomenologically by tuning the hopping parameters.

\section*{acknowledgments}
We thank C. Dean, A. Young, J.-H. Chu, D. Cobden, S. Chen, X. Liu, P. Kim, B. Lian, A. MacDonald, L. Levitov and L. Fu for helpful discussions. Technical support for the dilution refrigerator was provided by A. Manna and Z. Fei. This work was primarily supported by NSF MRSEC 1719797. X.X. acknowledges support from the Boeing Distinguished Professorship in Physics. X.X. and M.Y. acknowledge support from the State of Washington funded Clean Energy Institute. This work made use of a dilution refrigerator system which was provided by NSF DMR-1725221. Y.L. acknowledges the support of the China Scholarship Council. K.W. and T.T. were supported by the Elemental Strategy Initiative conducted by the MEXT, Japan and the CREST (JPMJCR15F3), JST.\\

\section*{Author contributions}
M.Y., X.X. and M.H. conceived the experiment. M.H. and Y.H.L. fabricated the devices, assisted by Y.L. M.H. performed the measurements, with assistance from Y.H.L. and Y.L. J.C. performed band structure calculation. K.W. and T.T. provided the bulk BN crystals. M.H., X.X., and M.Y. analyzed the data and wrote the paper with input from all authors.

\section*{Data Availability Statement}
The data that support the plots within this paper and other findings of this study are available from the corresponding author upon reasonable request.

\section*{Competing interests}
The authors declare no competing interests.

\section*{Additional Information}
Correspondence and requests for materials should be addressed to X.X. or M.Y.
%Supplementary information is available in the online version of the paper. Reprints and permission information is available online at www.nature.com/reprints. 

\section*{Supplementary Information}
Supplementary Sections 1-4 and Figs. 1-12.

\bibliographystyle{naturemag}
\bibliography{references}

\clearpage

%%%%%%%%%%%%%%%%%%%%%%%%%%%%%%%%%%%%%%%%%%%%%%%%%%%%%%%%%%%%%%%%%%%%%%%%%%%%%%%%%%%%%%%

\renewcommand{\thefigure}{S\arabic{figure}}
\renewcommand{\thesubsection}{S\arabic{subsection}}
\setcounter{secnumdepth}{2}
\renewcommand{\theequation}{S\arabic{equation}}
\renewcommand{\thetable}{S\arabic{table}}
%\subsubsectionfont{\normalfont\large\itshape\underline}
\setcounter{figure}{0} 
\setcounter{equation}{0}

\section*{Supplementary Information}

\subsection{Additional transport features in four tDBG devices with various twist angles}

\begin{figure*}[t]
\includegraphics[width=4 in]{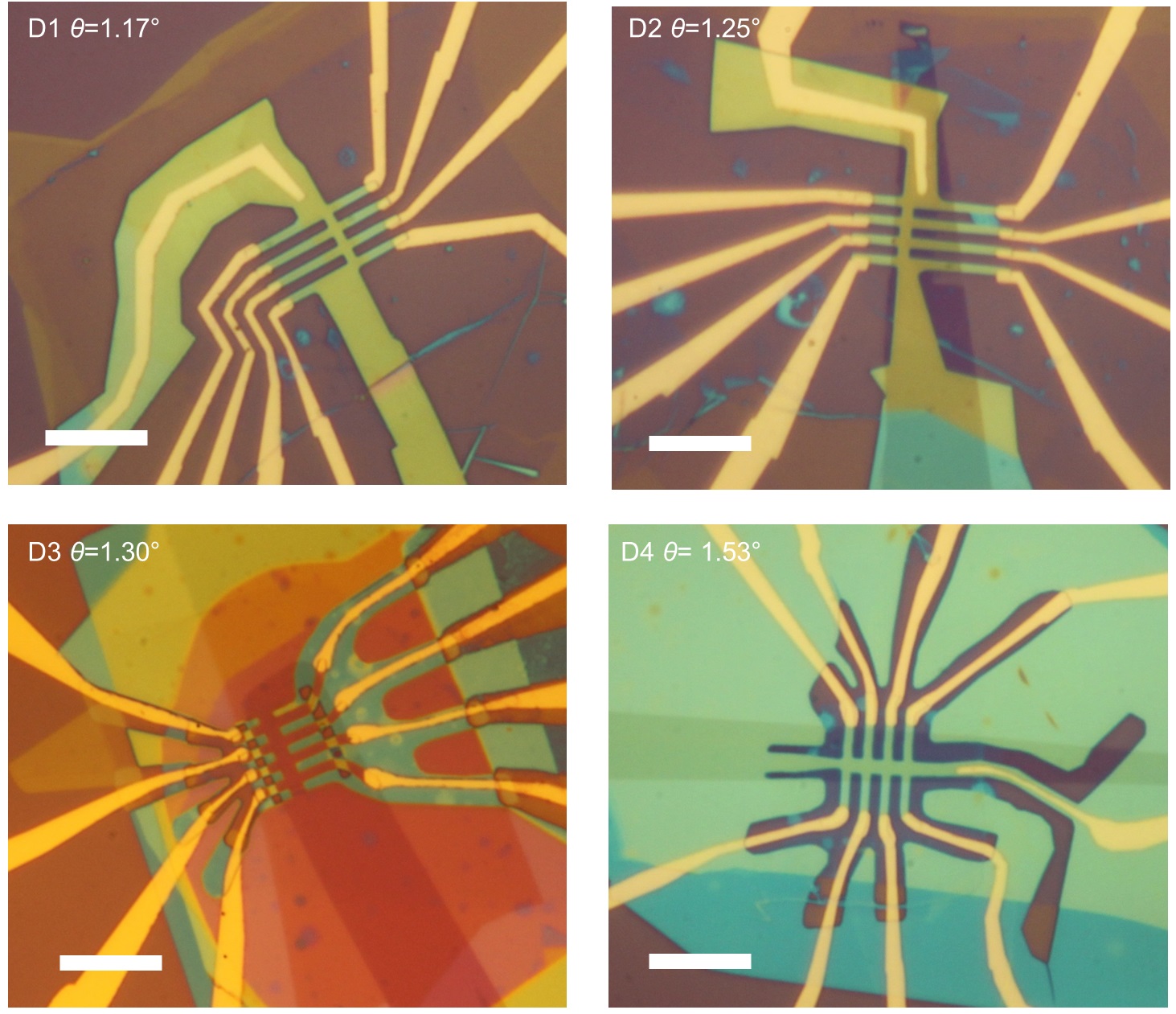} 
\caption{\textbf{Optical microscope images of the 4 tDBG devices.}
The twist angle of each device is denoted at the top left corner of each image. All scale bars are 10 $\mu$m.
}
\label{fig:S1}
\end{figure*}

\begin{figure*}[t]
\includegraphics[width=7 in]{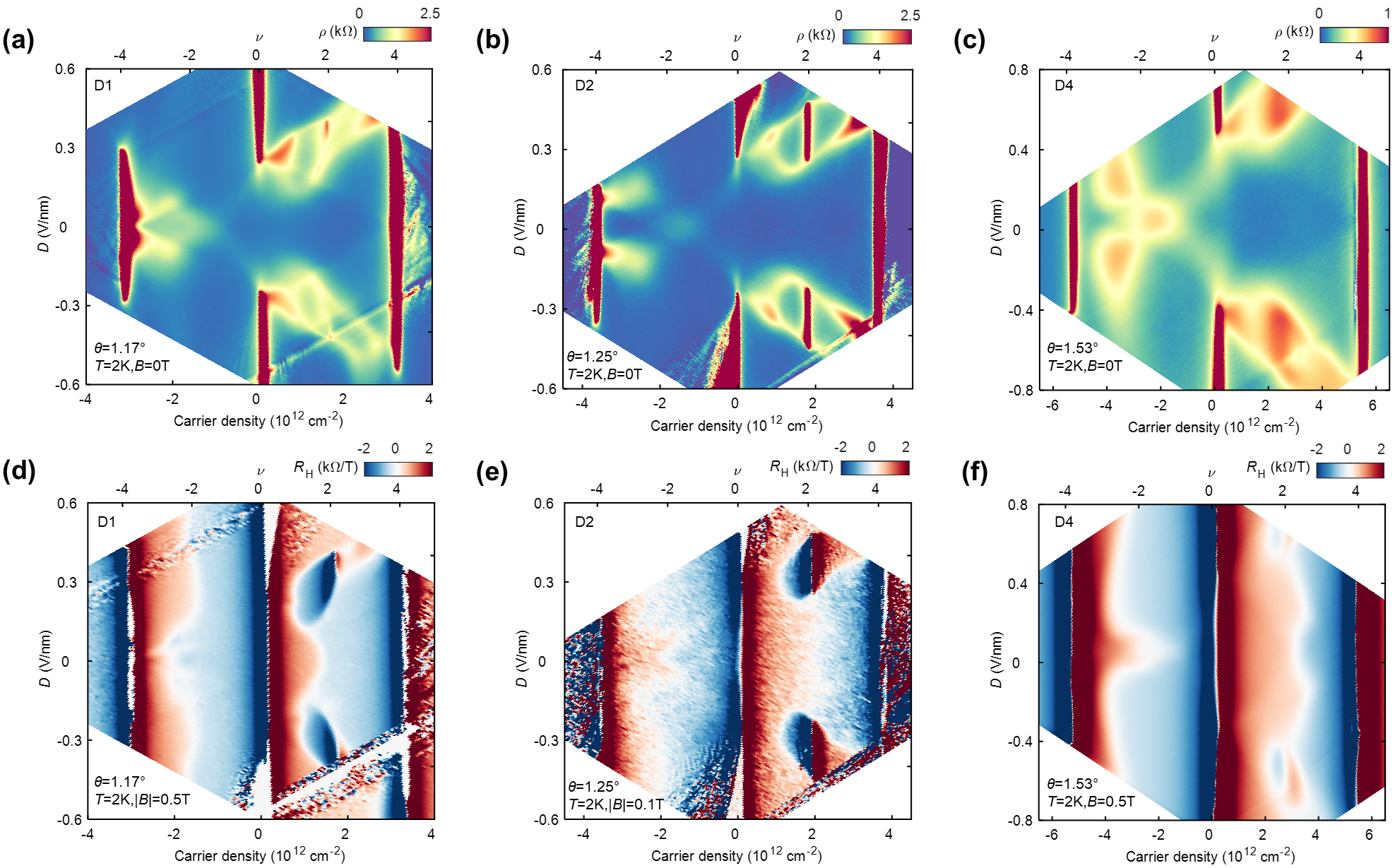} 
\caption{\textbf{Transport in tDBG at a variety of twist angles at $T = 2$~K.}
$\rho$ as a function of $n$ and $D$ in devices \textbf{a,} D1, \textbf{b,} D2, and \textbf{c,} D4 acquired at $B = 0$~T.
\textbf{d-f,} Corresponding $R_{H}$ for the same devices. The map in \textbf{d} is  antisymmetrized at $|B| = 0.5$~T, the map in \textbf{e} is antisymmetrized at $|B| = 0.1$~T, and the map in \textbf{f} is acquired at $B = +0.5$~T but not antisymmetrized. 
}
\label{fig:S2}
\end{figure*}

We investigate transport in a total of four tDBG devices with different twist angles, $\theta = 1.17^{\circ}$, 1.25$^{\circ}$, 1.30$^{\circ}$ and 1.53$^{\circ}$ (devices D1, D2, D3, and D4, respectively). Figure~\ref{fig:S1} shows optical microscope images of all four devices. Figure~\ref{fig:S2} shows $\rho$ and $R_{H}$ as a function of $n$ and $D$ for the three devices not discussed in the main text (\textit{i.e.} devices D1, D2, and D4).

We characterize the twist angle inhomogeneity by measuring two-terminal transport between all neighboring pairs of electrodes in our devices. Figure~\ref{fig:S10} shows this measurement for device D2. From the variation in carrier density corresponding to the $\nu = 2$ CI state, we infer a spread in twist angle of $\Delta\theta \approx 0.01^{\circ}$. Other devices in this study also exhibit relatively small twist angle inhomogeneity of $\Delta\theta \leq 0.03^{\circ}$.  

We observe qualitatively similar transport across all four devices (Fig.~\ref{fig:S2}), including a semimetal-to-semiconductor transition at $\nu = 0$ with increasing $|D|$, insulating states at $\nu = \pm4$, a cross-like feature for $\nu < 0$, and insulating or resistive states at $\nu = 2$ surrounded by halo features over a finite range of $|D|$. Insulating states at $\nu = \pm4$ exist over a smaller range of $|D|$ for smaller twist angles, consistent with tight-binding model calculations and with previous experimental reports~\cite{Shen2019,Liu2019,Cao2019b,Burg2019,Lee2019}.

The CI states are strongest in devices D2 and D3, weaker in device D1, and only weakly developed in device D4. In the absence of a magnetic field, incipient CI states at $\nu = 1$ and 3 are only observed in device D3 (Fig.~\ref{fig:S3}), however the base temperature for all other devices was limited to 2 K. Halo features are present in all four devices, suggesting that symmetry broken metallic states can be realized even when interactions are not sufficiently strong to completely gap the bands. Abrupt drops in $\rho(T)$ are observed within the halo regions of D1, D2, and D3. In addition to the data in Fig.~\ref{fig:2}b of the main text, Fig.~\ref{fig:S4} shows $\rho(T)$ for devices D1 and D2. The similarity of the transport within the halo regions of all four devices further confirms our interpretation that the correlated phase diagram is a consequence of spontaneous symmetry breaking.

The application of a small magnetic field $B \leq 0.5$~T appears to have only marginal effects on transport in our devices. Figure~\ref{fig:S5} shows $\rho$ and $R_H$ in device D3 at $T = 2$~K. In contrast to Fig.~\ref{fig:1}g of the main text, Fig.~\ref{fig:S5}b is acquired with ten times smaller $B = 0.05$~T. Although this results in smaller $R_H$, the primary features of the map are comparable to Fig.~\ref{fig:1}g. Additionally, we do not observe any signatures of hysteresis within the halo region as $B$ is swept (Fig.~\ref{fig:S7}, for device D2). So far, we have not observed any signatures of an anomalous Hall effect in our devices.

At higher temperatures, we observe substantial differences in transport between electron- and hole-type carriers at large $|D|$. Figure~\ref{fig:S11}a shows a map of $R_H$ in device D3, corresponding to the map of $\rho$ in Fig.~\ref{fig:1}b of the main text. A sign change is observed at the CNP only at large $|D|$, further demonstrating the semimetal-to-semiconducting crossover. Figure~\ref{fig:S11}b shows a map of $\rho$ as a function of $n$ and $T$ at $D = 0.27$~V/nm. There is a large difference in $\rho$ between electron- and hole-type doping at low $T$, but the difference becomes much less pronounced at higher $T$. We take this as further evidence that transport at low $T$ and large $D$ is mediated by isolated valence and conduction bands with different bandwidths, and hence different mobilities. As the temperature is raised, thermal activation mixes the bands. Finally, in all cases in which transport is nonlinear at base temperature, we observe linearity emerging at higher temperatures. Figure~\ref{fig:S12} shows an example corresponding to the purple curve in Fig.~\ref{fig:2} of the main text. 

\subsection{Evolution of transport features in an in-plane magnetic field}

Fig.~\ref{fig:S8} shows maps of $\rho$ and $R_{xy}$ for devices D2 and D3, analogous to those in Figs. 1f and g of the main text but with an in-plane magnetic field $B_{||} = 9$~T and a small ($< 0.5$~T) out-of-plane field, acquired at $T = 2$~K. CI states at $\nu = 1$, 2, and 3 are observed in both devices. Similar to the case of $B = 0$, we observe sign changes in $R_{xy}$ associated with the formation of each of these states, corresponding approximately to the halo regions in $\rho$. We also observe splitting of the van Hove singularity features for electron-type doping ($\nu < 0)$. We attribute this effect to Zeeman splitting of the valence band.

The application of $B_{||}$ also results higher base temperature resistance in all of the CI states. We extract the gap, $\Delta$, of the $\nu = 2$ CI state from thermal activation measurements in device D2 as a function of $D$ and $B_{||}$ (Fig.~\ref{fig:S6}a). The $\Delta(B_{||})$ appears to depend non-trivially on $D$. At small $D$, the gaps grow monotonically with $B_{||}$, whereas at larger $D$ the gaps first grow, then are reduced above a critical value of $B_{||}$. In contrast, at base temperature $\rho$ becomes more insulating for larger $B_{||}$ at all $D$, suggestive of spin-polarized states. However, $\rho(T)$ does not appear to exhibit simple thermal activation characteristics at large $D$, instead exhibiting kinks (Fig.~\ref{fig:S6}b). This unanticipated behavior may arise owing to a temperature-dependent many-body gap, and/or from a novel orbital effect~\cite{Lee2019} arising in an in-plane magnetic field owing to the multilayer structure of tDBG. Estimations of this effective orbital Zeeman term are of the same order of magnitude as the spin Zeeman term, complicating the ability to unambiguously dissociate the two contributions using the usual technique of simple tilting of the magnetic field.

\subsection{Quantum oscillations in tDBG} 

\begin{figure*}[t]
\includegraphics[width=6.5 in]{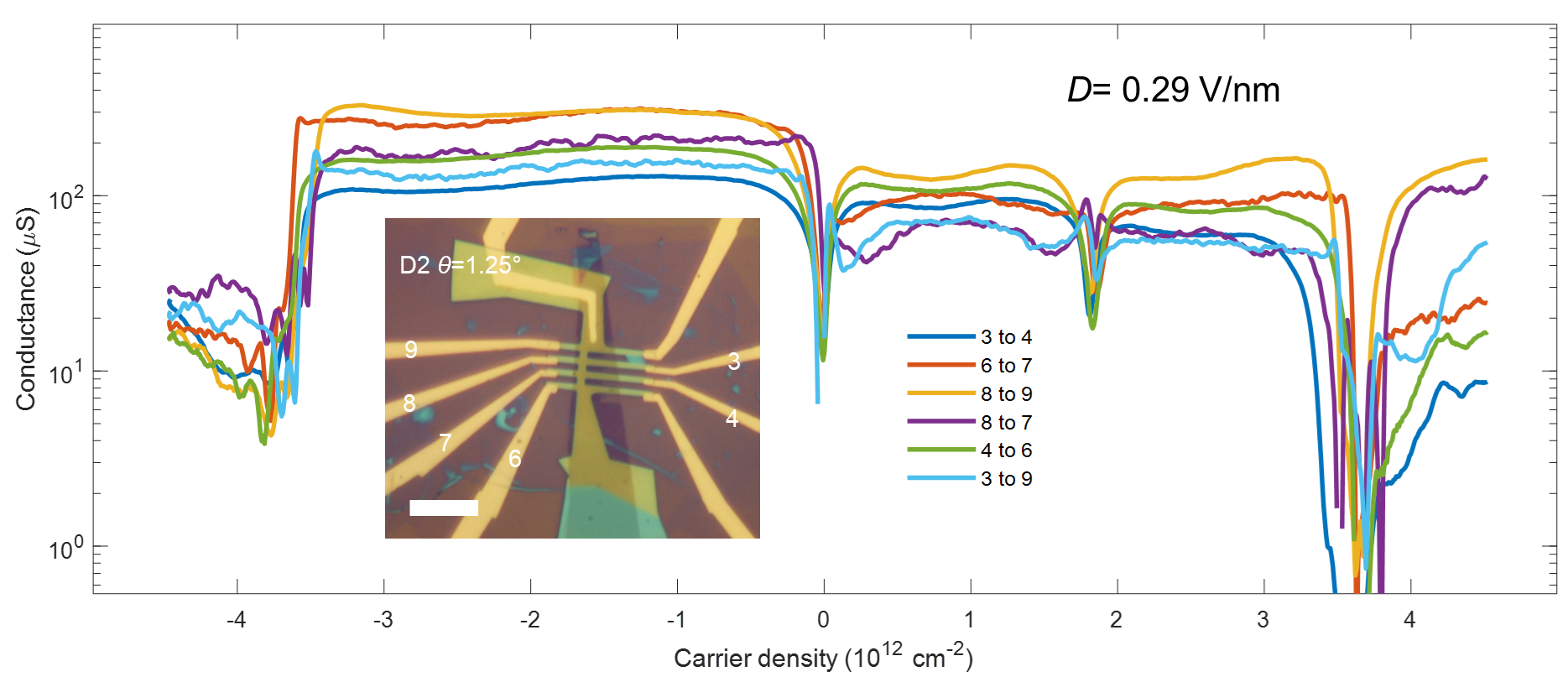} 
\caption{\textbf{Characterization of twist angle homogeneity in device D2.}
Two-terminal conductance measured as a function of $n$ at $D = 0.29$~V/nm. Transport is acquired between the two contacts denoted in the legend, matching the labels on the optical microscope image of the device in the inset. All other contacts are left floating during the measurement. The CI state at $\nu = 2$ is observed between every pair of contacts. The variation in $n$ of the CI state between different contacts pairs is $\sim 5 \times 10^{10}$ cm$^{-2}$, which corresponds to a spread in twist angle of $\Delta\theta \approx 0.01^{\circ}$.}
\label{fig:S10}
\end{figure*}

\begin{figure*}[t]
\includegraphics[width=6.5 in]{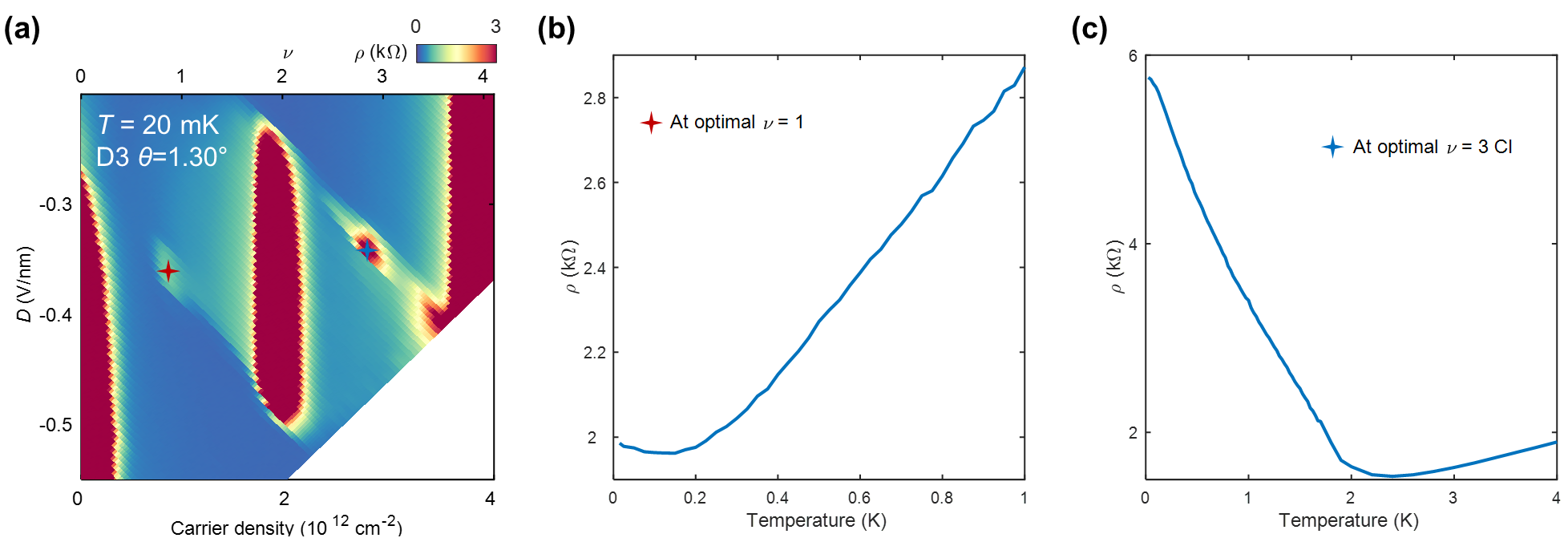} 
\caption{\textbf{Transport of the CI states at $\nu = 1$ and $3$.}
\textbf{a,} $\rho$ map surrounding the CI states in device D3 at $T = 20$~mK. 
\textbf{b,} $\rho(T)$ acquired where the $\nu = 1$ CI state is most resistive. A weak metal-insulator transition is observed at $T \approx 150$~mK.
\textbf{c,} $\rho(T)$ of the $\nu = 3$ CI state, exhibiting a metal-insulator transition at $T \approx 2.5$~K. 
}
\label{fig:S3}
\end{figure*}

\begin{figure*}[t]
\includegraphics[width=4.5 in]{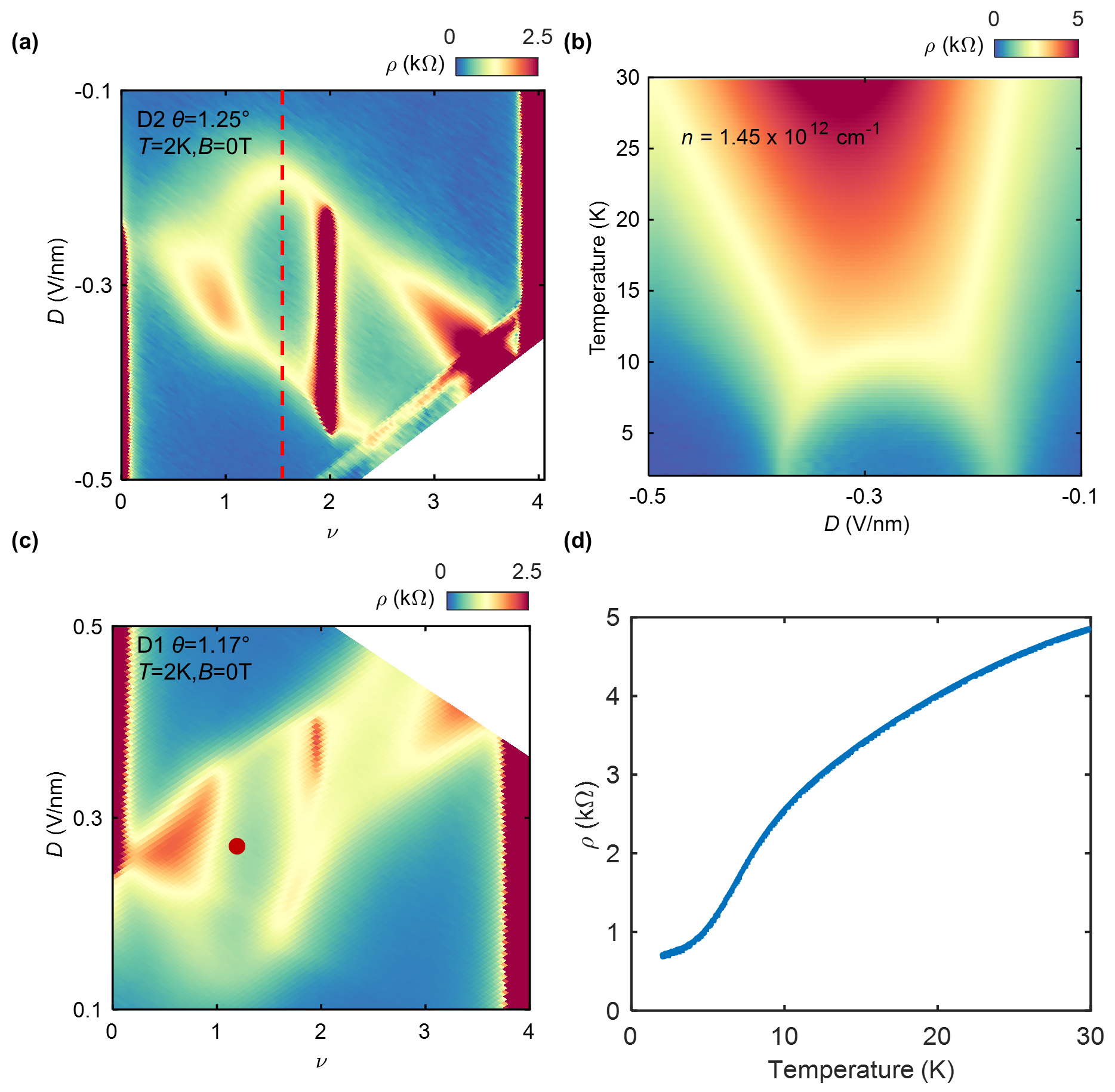} 
\caption{\textbf{Transport near the halo regions in additional devices.}
\textbf{a,} $\rho$ map in device D2 at $T = 2$~K, exhibiting a CI state at $\nu = 2$ and a halo feature.
\textbf{b,} $\rho(T)$ as a function of $D$ at fixed $\nu$ (acquired along the dashed red line in \textbf{a}. An arch-like feature is observed, similar to the behavior of device D3 shown in Fig.~\ref{fig:2}a of the main text.
\textbf{c,} $\rho$ map in device D1 at $T = 2$~K. A very weak CI state is observed at $\nu = 2$, as well as a distorted halo feature.
\textbf{d,} $\rho(T)$ acquired at the point marked by the red circle in \textbf{c}. Despite weaker correlations, an abrupt drop in $\rho(T)$ is still observed to accompany the formation of the symmetry-broken halo region.
}
\label{fig:S4}
\end{figure*}

\begin{figure*}[t]
\includegraphics[width=5 in]{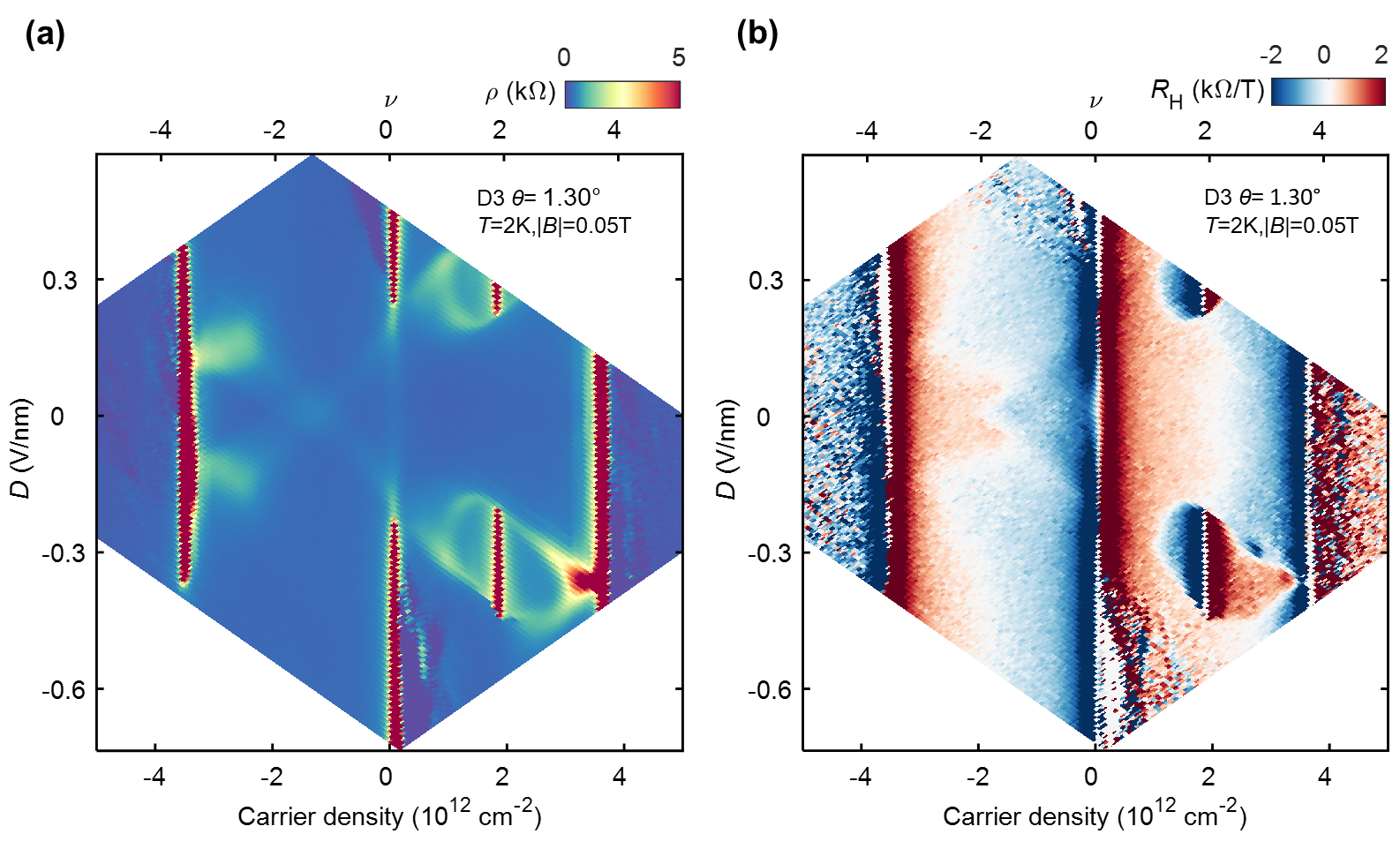} 
\caption{\textbf{Transport in device D3 at $T = 2$~K.}
\textbf{a,} Map of $\rho$ comparable to Fig.~\ref{fig:1}f of the main text. The halo and cross features are broadened at higher temperature.
\textbf{b,} $R_H$ map antisymmetrized at $B = 0.05$~T. The overall features are similar to Fig.~\ref{fig:1}g of the main text, suggesting that increasing the magnetic field to 0.5~T does not substantially modify the transport. 
}
\label{fig:S5}
\end{figure*}

\begin{figure*}[t]
\includegraphics[width=4 in]{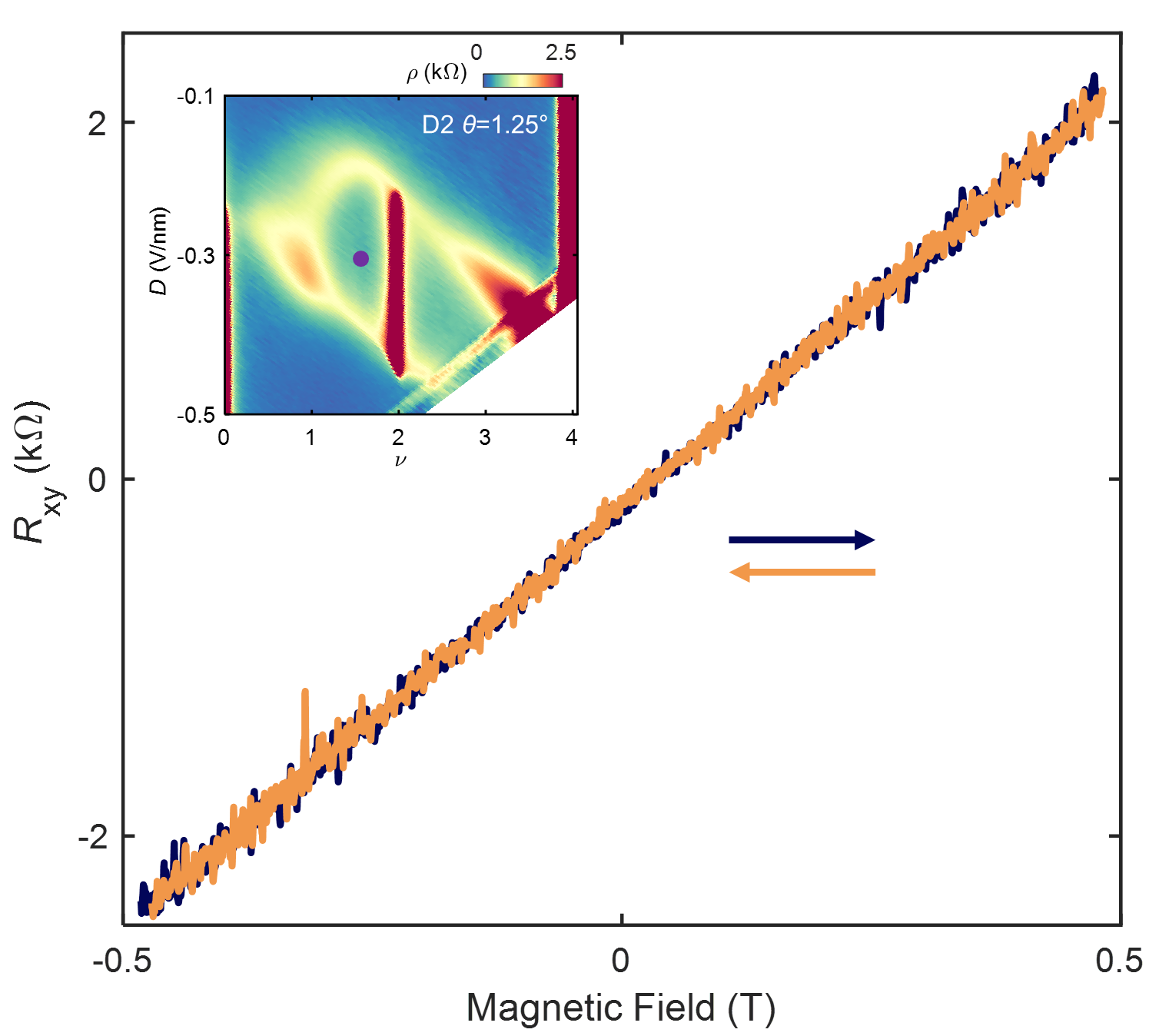} 
\caption{\textbf{$R_{xy}(B)$ within the halo region in device D2 at $T = 2$~K.}
$R_{xy}$ appears to be independent of the sweeping direction of the magnetic field, pointing to the absence of an anomalous Hall effect. The data is acquired at the point marked by the purple circle in the inset. No signature of hysteresis was observed at any of the points that were tested, both inside and nearby the halo.
}
\label{fig:S7}
\end{figure*}

\begin{figure*}[t]
\includegraphics[width=5.5 in]{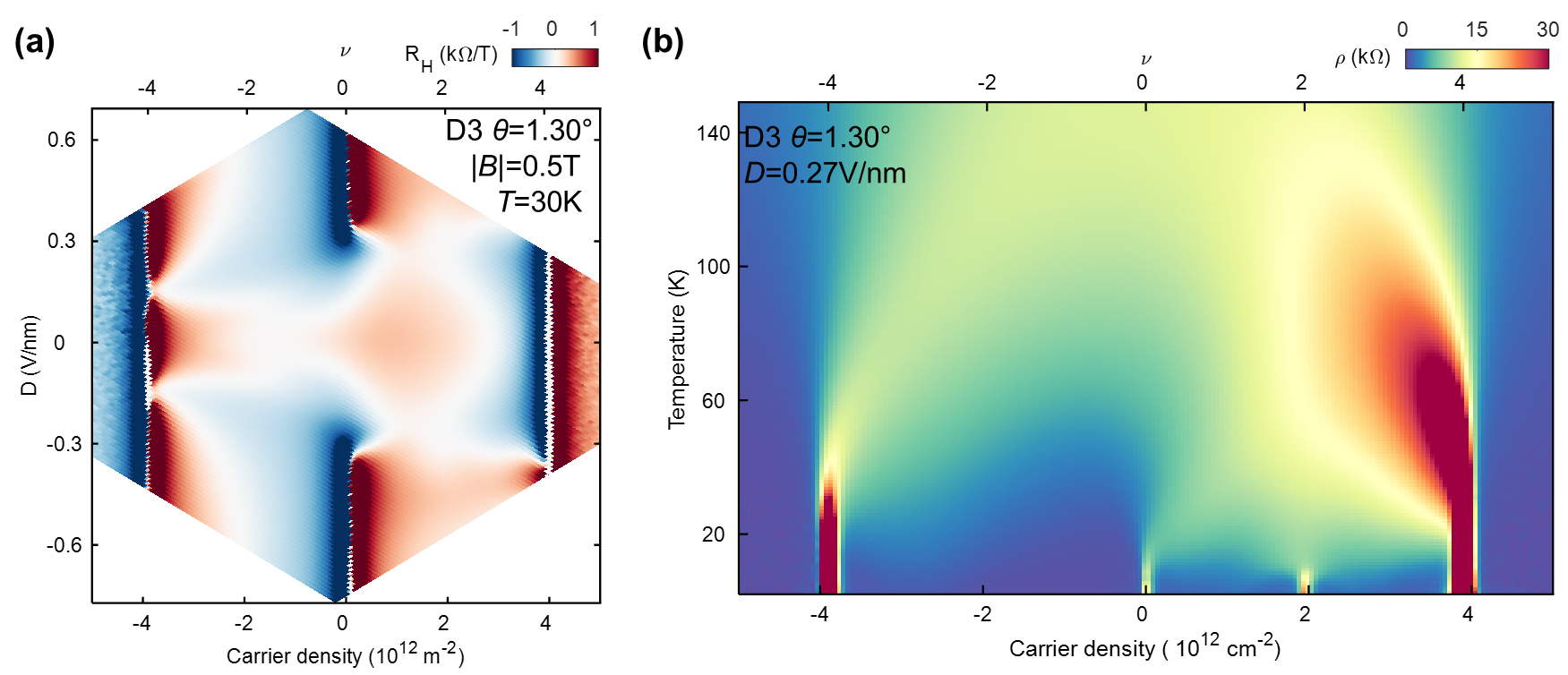} 
\caption{\textbf{High temperature transport in device D3.}
\textbf{a,} $R_H$ map at $|B| = 0.5$~T and $T = 30$~K, associated with the $\rho$ map in Fig.~\ref{fig:1}b of the main text. Semi-metallic behavior is observed at small $|D|$, whereas semiconducting behavior emerges at larger $|D|$ marked by a sign reversal at the CNP. 
\textbf{b,} $\rho$ at a function of $n$ and $T$ at $D = 0.27$~V/nm. Large differences in $\rho$ between electron- and hole-type doping are evident at low temperature, but become more similar at high temperature owing to thermal activation between the bands.
}
\label{fig:S11}
\end{figure*}

\begin{figure*}[t]
\includegraphics[width=3 in]{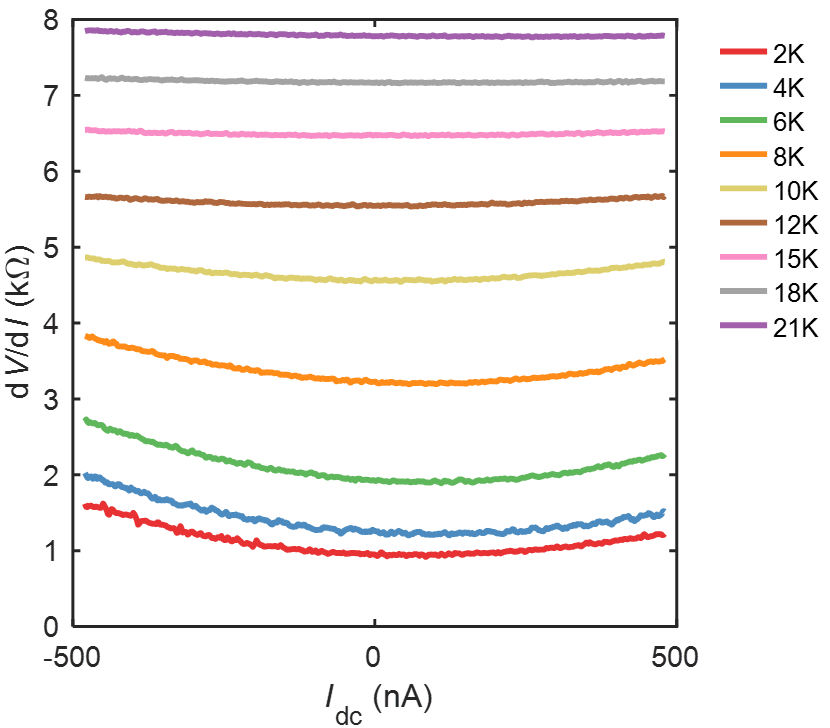} 
\caption{\textbf{Temperature dependence of d$V$/d$I$ in device D3 .}
Data is acquired at $(n,D)$ corresponding to the purple curves in Fig.~\ref{fig:2} of the main text.
}
\label{fig:S12}
\end{figure*}

\begin{figure*}[t]
\includegraphics[width=5 in]{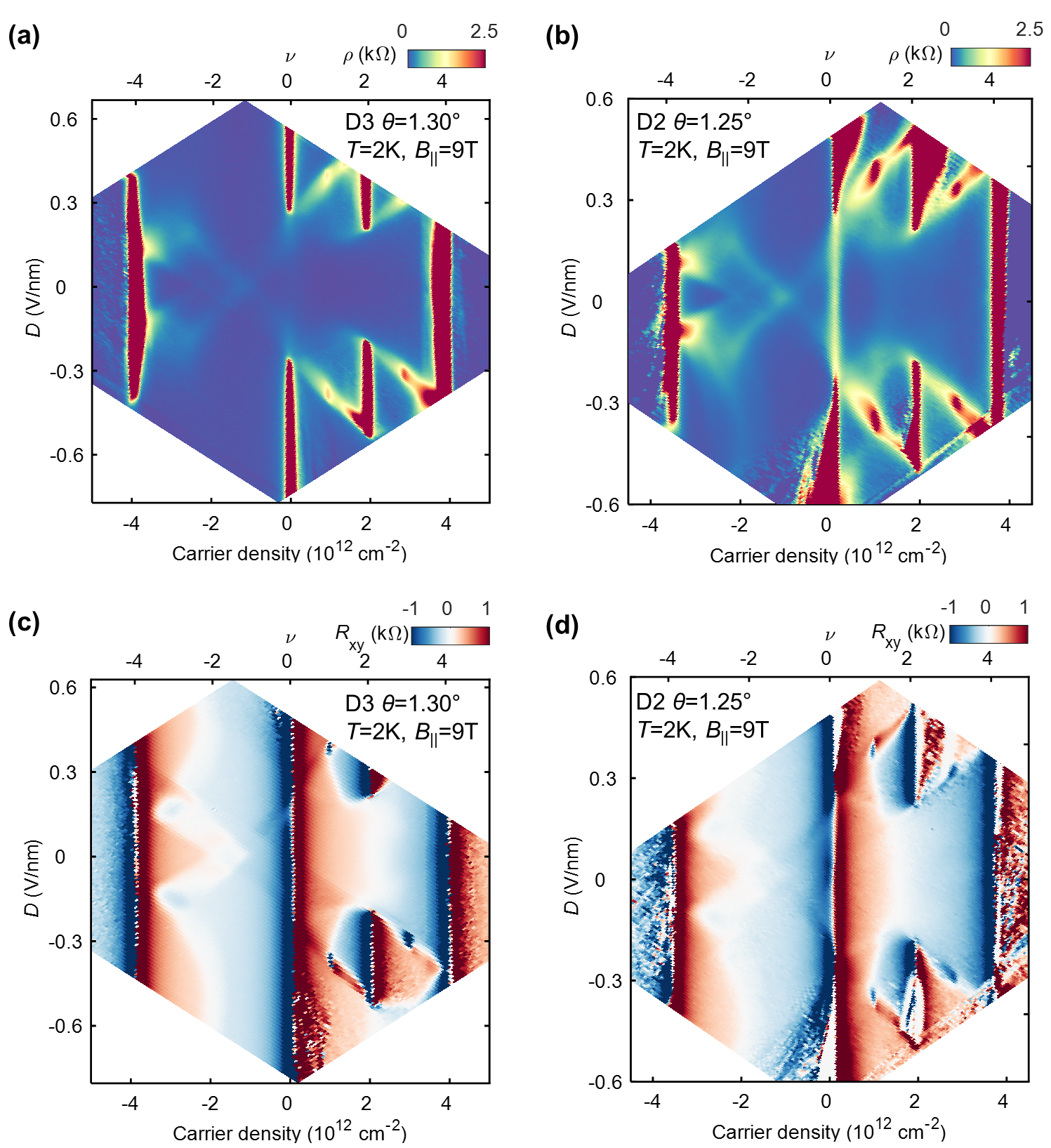} 
\caption{\textbf{Transport at $B_{||} = 9$~T in devices D2 and D3.}
Maps of $\rho$ at $T = 2$~K in devices \textbf{a,} D3 and \textbf{b,} D2, along with corresponding $R_{xy}$ in \textbf{c} and \textbf{d}. A small out-of-plane $B$ is also added.
}
\label{fig:S8}
\end{figure*}

\begin{figure*}[t]
\includegraphics[width=6.5 in]{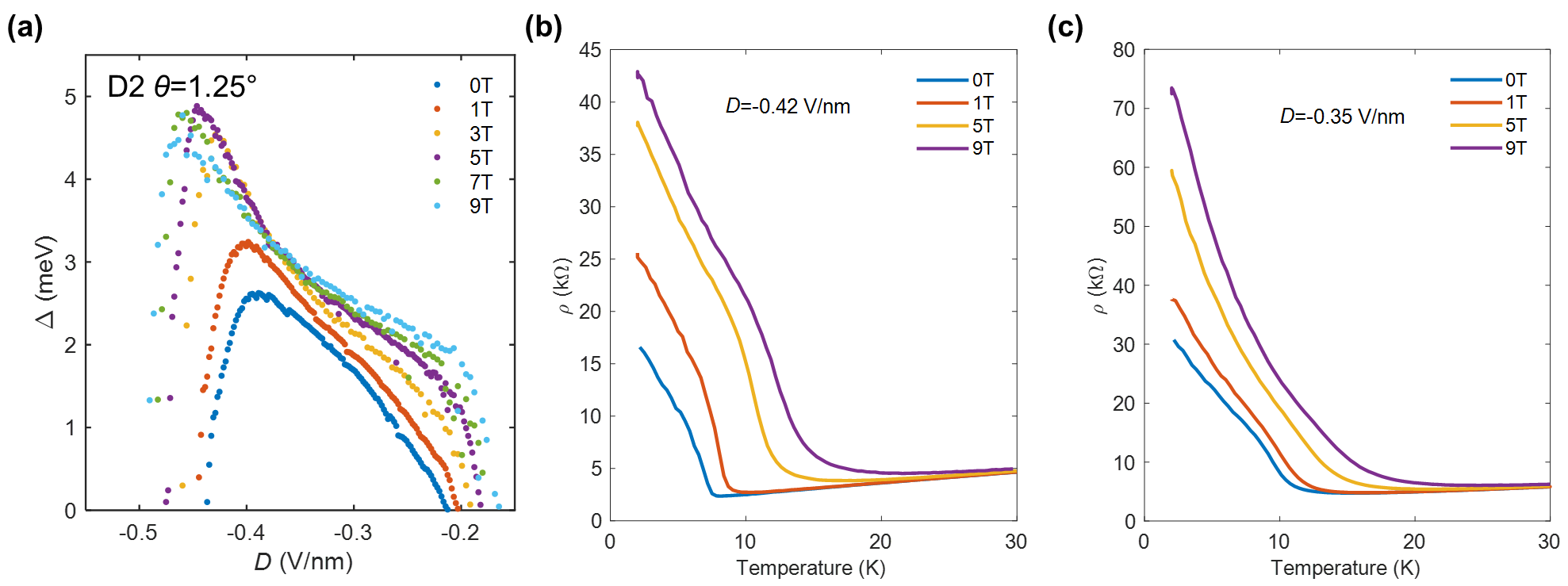} 
\caption{\textbf{Energy gaps of the $\nu = 2$ CI state with $B_{||}$ in device D2.}
\textbf{a,} Gaps measured by thermal activation, with examples of corresponding $\rho(T)$ in \textbf{b,} at $D = -0.42$~V/nm and \textbf{c,} at $D = -0.35$~V/nm. Kinks develop in $\rho(T)$ for larger $|D|$, indicating a departure from simple activated transport.
}
\label{fig:S6}
\end{figure*}

\begin{figure*}[t]
\includegraphics[width=5.5 in]{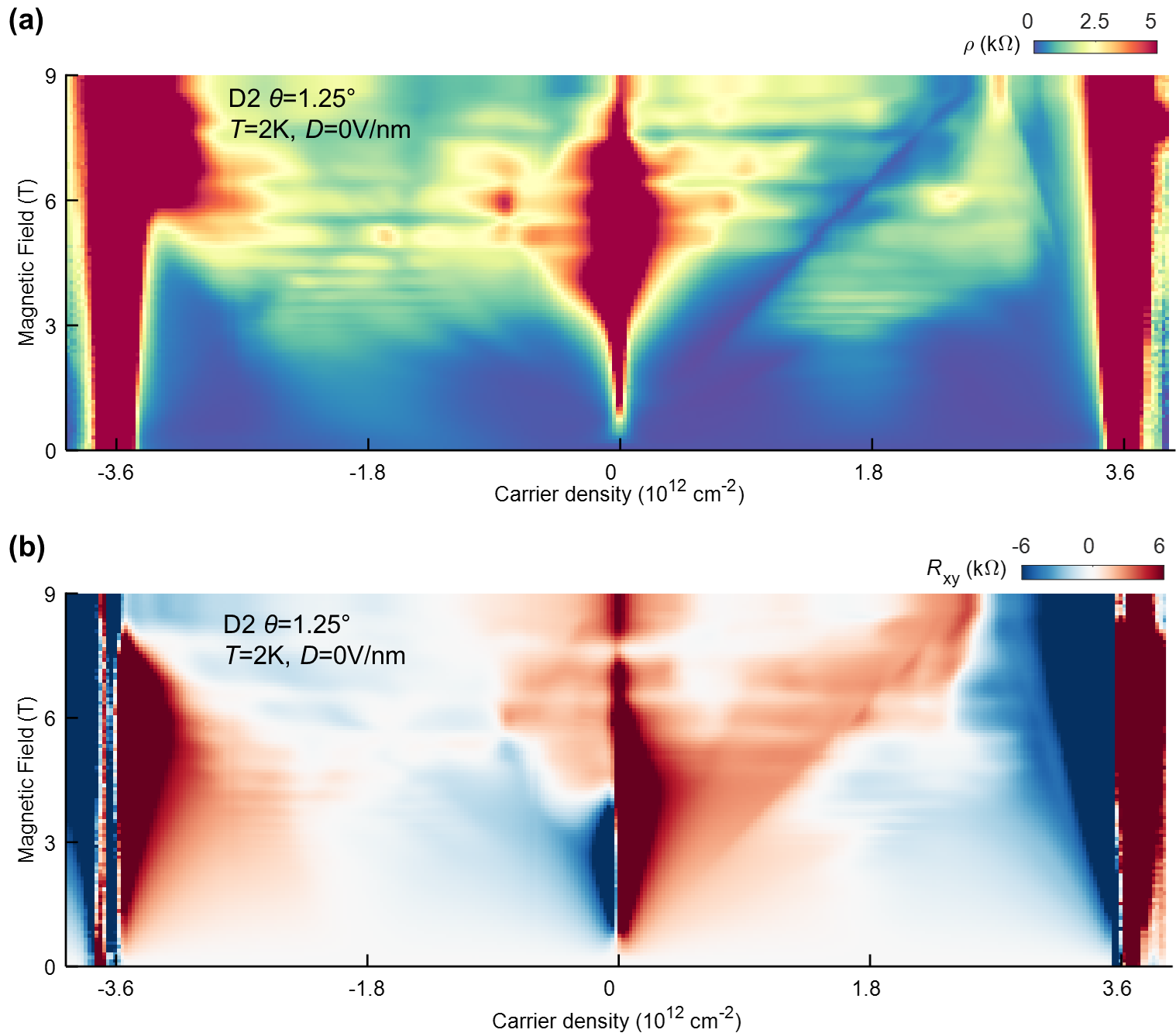} 
\caption{\textbf{Quantum oscillations in device D2 at $T = 2$~K.}
Landau fan diagram of \textbf{a,} $R_{xx}$ and \textbf{b,} $R_{xy}$ at $D = 0$~V/nm. The dominant quantum oscillation on the electron side corresponds to $\nu = +6$.
}
\label{fig:S9}
\end{figure*}

Fig.~\ref{fig:S9} shows Landau fan diagrams for device D2 acquired at $D = 0$~V/nm. Despite the apparent high homogeneity of the twist angle in our devices (Fig.~\ref{fig:S10}), quantum oscillations are rather poorly developed in all four devices. In particular, we typically only observe main-sequence quantum Hall states, and do not observe well-developed symmetry-broken states. Although this may be associated with the small cyclotron gaps of the flat bands and/or twist angle inhomogeneity, we do not currently have a complete understanding of this observation. Nevertheless, in general we observe quantum oscillations with filling factor sequence $+6$, $+12$, etc. for electron-type doping. For hole-type doping, within our resolution quantum oscillations appear to be consistent with filling factor sequence $-4$, $-8$, $-12$, etc. Upon the emergence of correlated states at larger $D$, quantum oscillations become too poorly resolved to confidently identify their filling factor sequence. 

The 6-fold degeneracy for electron-type doping is distinctively different from the typical 4-fold spin-valley degeneracy in bilayer graphene. Tight-binding model calculations of tDBG (Fig.~\ref{fig:1}c of the main text) suggest a complicated band structure, in which there are multiple Fermi surfaces within the moir\'e-Brillouin zone. Although there remain four degenerate copies of each flat band, these additional Fermi surface pockets may contribute additional degeneracies to measurements of the quantum oscillations. A combination of higher quality devices and careful theoretical analysis will be necessary to resolve this unexpected observation.

Despite the poorly resolved quantum oscillations, our devices always exhibit strong Brown-Zak oscillations~\cite{Kumar2017} reflective of the Hofstadter’s butterfly spectrum (\textit{i.e.} the horizontal stripes in Fig.~\ref{fig:S9}). These oscillations arise at magnetic fields corresponding to rational flux filling of the moir\'e unit cell. Fitting this sequence of features provides an independent confirmation of the extracted twist angle for each device.

\subsection{Comparison with tBLG} 

Our observations in tDBG contrast previous results in tBLG in a number of ways. First, tBLG exhibits numerous clear signatures of superconductivity: rapid drops to immeasurably small $\rho$ below a critical temperature, well-defined critical currents in $I$-$V$ measurements, and signatures of Fraunhofer-like phase coherent transport in a magnetic field owing to sample inhomogeneity. Second, recent local spectroscopy measurements in tBLG have revealed a sequence of abrupt symmetry breakings in which the initial four-fold degeneracy of each flat band is reduced by one each time the chemical potential crosses integer $\nu$~\cite{Wong2019,Zondiner2019}. Correspondingly, transport measurements exhibit abrupt resets of $n_H$ to zero only upon overdoping of a well-developed CI state, and generally do not come with an associated sign change~\cite{Cao2018b,Yankowitz2019,Lu2019,Stepanov2019,Saito2019}. This behavior contrasts our observations in tDBG and previous observations in other semiconducting flat band systems, where the sign of $n_H$ reverses and a symmetry is broken even when the chemical potential is at substantial underdoping of a CI state. Notably, similar sign changes in $n_H$ are also observed surrounding $\nu = 2$ in tBLG aligned with BN~\cite{Serlin2019}, in which $C_2$ rotation symmetry is broken and a single-particle gap opens at $\nu = 0$. Together, this may point to the critical role of $C_2 T$ symmetry in determining the correlated phase diagram of tBLG, however more experiments will be necessary to adequately address this point. 

%%%%%%%%%%%%%%%%%%%%%%%%%%%%%%%%%%%%%%%%%%%%%%%%%%%%%%%%%%%%%%%%%%%%%%
%%%%%%%%%%%%%%%%%%%%%%%%%%%%%%%%%%%%%%%%%%%%%%%%%%%%%%%%%%%%%%%%%%%%%%

%%%%%%%%%%%%%%%%%%%%%%%%%%%%%%%%%%%%%%%%%%%%%%%%%%%%%%%%%%%%%%%%%%%%%%
%%%%%%%%%%%%%%%%%%%%%%%%%%%%%%%%%%%%%%%%%%%%%%%%%%%%%%%%%%%%%%%%%%%%%%

%%%%%%%%%%%%%%%%%%%%%%%%%%%%%%%%%%%%%%%%%%%%%%%%%%%%%%%%%%%%%%%%%%%%%%
%%%%%%%%%%%%%%%%%%%%%%%%%%%%%%%%%%%%%%%%%%%%%%%%%%%%%%%%%%%%%%%%%%%%%%

%%%%%%%%%%%%%%%%%%%%%%%%%%%%%%%%%%%%%%%%%%%%%%%%%%%%%%%%%%%%%%%%%%%%%%
%%%%%%%%%%%%%%%%%%%%%%%%%%%%%%%%%%%%%%%%%%%%%%%%%%%%%%%%%%%%%%%%%%%%%%

%%%%%%%%%%%%%%%%%%%%%%%%%%%%%%%%%%%%%%%%%%%%%%%%%%%%%%%%%%%%%%%%%%%%%%
%%%%%%%%%%%%%%%%%%%%%%%%%%%%%%%%%%%%%%%%%%%%%%%%%%%%%%%%%%%%%%%%%%%%%%

%%%%%%%%%%%%%%%%%%%%%%%%%%%%%%%%%%%%%%%%%%%%%%%%%%%%%%%%%%%%%%%%%%%%%%
%%%%%%%%%%%%%%%%%%%%%%%%%%%%%%%%%%%%%%%%%%%%%%%%%%%%%%%%%%%%%%%%%%%%%%

%%%%%%%%%%%%%%%%%%%%%%%%%%%%%%%%%%%%%%%%%%%%%%%%%%%%%%%%%%%%%%%%%%%%%%
%%%%%%%%%%%%%%%%%%%%%%%%%%%%%%%%%%%%%%%%%%%%%%%%%%%%%%%%%%%%%%%%%%%%%%

%%%%%%%%%%%%%%%%%%%%%%%%%%%%%%%%%%%%%%%%%%%%%%%%%%%%%%%%%%%%%%%%%%%%%%
%%%%%%%%%%%%%%%%%%%%%%%%%%%%%%%%%%%%%%%%%%%%%%%%%%%%%%%%%%%%%%%%%%%%%

%%%%%%%%%%%%%%%%%%%%%%%%%%%%%%%%%%%%%%%%%%%%%%%%%%%%%%%%%%%%%%%%%%%%%%
%%%%%%%%%%%%%%%%%%%%%%%%%%%%%%%%%%%%%%%%%%%%%%%%%%%%%%%%%%%%%%%%%%%%%

%%%%%%%%%%%%%%%%%%%%%%%%%%%%%%%%%%%%%%%%%%%%%%%%%%%%%%%%%%%%%%%%%%%%%%
%%%%%%%%%%%%%%%%%%%%%%%%%%%%%%%%%%%%%%%%%%%%%%%%%%%%%%%%%%%%%%%%%%%%%%

%%%%%%%%%%%%%%%%%%%%%%%%%%%%%%%%%%%%%%%%%%%%%%%%%%%%%%%%%%%%%%%%%%%%%%
%%%%%%%%%%%%%%%%%%%%%%%%%%%%%%%%%%%%%%%%%%%%%%%%%%%%%%%%%%%%%%%%%%%%%%

%%%%%%%%%%%%%%%%%%%%%%%%%%%%%%%%%%%%%%%%%%%%%%%%%%%%%%%%%%%%%%%%%%%%%%
%%%%%%%%%%%%%%%%%%%%%%%%%%%%%%%%%%%%%%%%%%%%%%%%%%%%%%%%%%%%%%%%%%%%%

\end{document}